\documentclass{eptcs}

\usepackage[heading-base={2},section-num={False},bib-label={True}]{madoko2}
\usepackage{times}
\usepackage[T1]{fontenc}

    \IfFileExists{luximono.sty}{\usepackage[scaled=0.81]{luximono}}{\usepackage[scaled=0.81]{beramono}}
    \def\authorrunning{J. Koenig and K.R.M. Leino} 
    \renewcommand\mdauthorrunning[1]{}

\begin{document}

\mdxtitleblockstart{}
\mdxtitle{\mdline{30}Programming Language Features for Refinement}
\mdxauthorstart{}
\mdxauthorname{\mdline{35}Jason Koenig}

\mdxauthoraddress{\mdline{38}Stanford University}

\mdxauthoremail{\mdline{41}jrkoenig@stanford.edu}
\mdxauthorend\mdxauthorstart{}
\mdxauthorname{\mdline{46}K. Rustan M. Leino}

\mdxauthoraddress{\mdline{49}Microsoft Research}

\mdxauthoremail{\mdline{52}leino@microsoft.com}
\mdxauthorend\mdtitleauthorrunning{}{}\mdxtitleblockend

\begin{abstract}

\noindent\mdline{33}Algorithmic and data refinement are well studied topics that provide a
mathematically rigorous approach to gradually introducing details in
the implementation of software.  Program refinements are performed in
the context of some programming language, but mainstream languages
lack features for recording the sequence of refinement steps in the
program text.  To experiment with the combination of refinement,
automated verification, and language design, refinement features have
been added to the verification-aware programming language Dafny.  This
paper describes those features and reflects on some initial usage
thereof.
\end{abstract}

\section{\mdline{45}0.\hspace*{0.5em}\mdline{45}Introduction}\label{sec-introduction}

\noindent\mdline{47}Two major problems faced by software engineers are the development of
software and the maintenance of software.  In addition to fixing bugs,
maintenance involves adapting the software to new or previously
underappreciated scenarios, for example, using new APIs, supporting
new hardware, or improving the performance.  Software version control
systems track the history of software changes, but older versions
typically do not play any significant role in understanding or
evolving the software further.  For example, when a simple but
inefficient data structure is replaced by a more efficient one,
the program edits are destructive.  Consequently, understanding the
new code may be significantly more difficult than understanding the
initial version, because the source code will only show how the more
complicated data structure is used.

\mdline{61}The initial development of the software may proceed in a similar way,
whereby a software engineer first implements the basic functionality
and then extends it with additional functionality or more advanced
behaviors.  For example, the development of a library that provides
binary decision diagrams (BDDs) may proceed as follows: the initial
version may use a simple data structure; then, reductions are
implemented; to facilitate quicker look-ups, hash-consing is added;
caches are added to speed up commonly occurring operations; the
garbage collection provided by the programming language is replaced by
a custom allocator and collector that are specific to the needs of the
BDD library; the functionality is extended to allow setting the
variable ordering; watch dogs are added to monitor how the variable
ordering is affecting performance; a system for automatically changing
the variable ordering dynamically is added.  Much before these steps
have all been added, the software has reached considerable complexity
and has become difficult to understand and costly to maintain.

\mdline{78}If the design of a piece of software were explained from one software
engineer to another, the explanations would surely be staged to
explain the subsequent layers of complexity gradually.  In this paper,
we consider how a \mdline{81}\emph{programming language}\mdline{81} can give software engineers
the ability to write the source code in logical stages, in the same
way that it may be explained in person.  In particular, we describe
our design of \mdline{84}\emph{refinement}\mdline{84} features in the verification-aware
programming language Dafny\mdline{85}~[\mdcite{leino:dafny:lpar16}{16}]\mdline{85}.

\mdline{87}Stepwise program refinement, including data refinement, has been
studied a great deal in the last several decades, see, e.g.,
\mdline{89}[\mdcite{abrial:eventb:book}{0}, \mdcite{backvonwright:book}{3}, \mdcite{morgan:book}{23}, \mdcite{kiv:overview}{26}]\mdline{89}.
It provides a mathematical framework for gradually introducing
complexity into a
design.  It has been implemented in software construction and modeling
tools, like KIV\mdline{93}~[\mdcite{kiv:overview}{26}]\mdline{93}, Atelier B, and Rodin\mdline{93}~[\mdcite{rodintoolset}{1}]\mdline{93}.  However, the input to these
tools take a larger departure from today\mdline{94}'\mdline{94}s programming languages than
we would like.  From the language design perspective, we are looking
for something more in the spirit of the Transform
\mdline{97}[\mdcite{griesprins:encapsulation}{8}, \mdcite{griesvolpano:transform}{9}]\mdline{97} or SETL\mdline{97}~[\mdcite{setl}{27}]\mdline{97},
but with tool support for both compilation and reasoning.  As we report in
this paper, we have found it difficult to design a usable set of
features in the programming language.  We are not ready to give up,
however.  Instead, we hope that our mixed
experiences will inspire improved designs in the future.

\mdline{104}We describe our design goals in Section\mdline{104}~\mdref{sec-design-goals}{1}\mdline{104}.
In Section\mdline{105}~\mdref{sec-refinement-in-dafny}{2}\mdline{105}, we describe Dafny\mdline{105}'\mdline{105}s refinement
features and illustrate these with small examples.
Refining an instance of a simple class into an aggregate object that
uses new instances of library-defined classes is difficult
\mdline{109}[\mdcite{filipovic:seplogicrefinement}{6}, \mdcite{leinoyessenov:chalicerefinement}{20}]\mdline{109}.
In Section\mdline{110}~\mdref{sec-classes}{3}\mdline{110}, we show how this is done in Dafny.
In the last sections of the paper, we reflect on our experience,
compare with related work, and conclude.

\section{\mdline{114}1.\hspace*{0.5em}\mdline{114}Design Goals}\label{sec-design-goals}

\noindent\mdline{116}Our view is that the programming language is a software engineer\mdline{116}'\mdline{116}s
most important tool.  Therefore, we think it is important to try to
capture more of the design of a program into the program text itself.
A program can use the constructs in a language to aid in making a
design understandable, which is important both for development and
maintenance.

\mdline{123}A central pedagogical principle lies in presenting
details at the right time, and this principle is manifested in many
well-known programming facilities.
Among these, \mdline{126}\emph{procedural abstraction}\mdline{126}\textemdash{}\mdline{126}whereby computational
descriptions are divided into named, reusable routines\mdline{127}\textemdash{}\mdline{127}is perhaps
the most universal.  \mdline{128}\emph{(Interface and implementation) modules}\mdline{128}
provide another way to hide details, for example as the \mdline{129}\textquotedblleft{}one secret
per module\textquotedblright{}\mdline{130} guideline enunciated by Parnas\mdline{130}~[\mdcite{parnas:secret}{25}]\mdline{130}.  In object-oriented
software, \mdline{131}\emph{subclassing}\mdline{131} gives a way to collect common behavior and to
customize details in class-specific ways.  In functional programming,
\mdline{133}\emph{type parametricity}\mdline{133} gives a way to operate over data without needing
to be concerned with the specifics of the data, thus abstracting over
the details.  Cross-cutting details can also be introduced using
\mdline{136}\emph{aspects}\mdline{136}, which give a whole-program way to customize behavior
\mdline{137}[\mdcite{aspects:invitedtalk-ecoop1997}{14}]\mdline{137}.
We are hoping for refinement features that give yet another way to
stage the complexity of a program.  Whereas procedural abstraction
allows layering of the call graph, refinement features aim to layer
the logical complexity.

\mdline{143}Our design goals are to provide:

\begin{itemize}

\item{}
\mdline{145}Programming in stages.  We want the refinement features to allow
logical, gradual introduction of details.  We also want the result
to be easier to understand than what alternative constructs provide
today.

\item{}
\mdline{150}A program structuring device.  While it seems desirable for modules
to provide strict information-hiding barriers, this is not usually a
strong concern in procedural abstraction.  Procedures internal to a
module frequently factor out behavior without going as far as making
sure callers and callees are entirely decoupled.  For example, a
change in the caller may require a change in the callee, and vice
versa.  It may be helpful to think of this as the \mdline{156}\emph{one-developer
view}\mdline{157}, where the one software developer is in control of
both sides of the procedural abstraction boundary.  Our aim is for
the refinement features to have such a one-developer view.  That is,
we do not see a refinement boundary as a boundary that must
support all sorts of uses.  Instead, a refinement may be introduced
as a program structuring device that just helps organize the program
into logically staged pieces.  A developer will not be shamed when
making changes to the software that require changes to other sides
of refinement boundaries.  In particular, we will allow an initial
design to \mdline{166}\emph{anticipate}\mdline{166} further refinements.  For example, this makes
it okay for the program to contain \mdline{167}\textquotedblleft{}shims\textquotedblright{}\mdline{167} or \mdline{167}\textquotedblleft{}refinement points\textquotedblright{}\mdline{167}
that are to be filled in or referenced later.

\item{}
\mdline{170}Program-like constructs.  By considering refinement in the
programming language, we are doing something that is different from
mainstream programming languages.  However, we do not want to stray
too far\mdline{173}\textemdash{}\mdline{173}we want the result to still look, more or less, like
mainstream programs today.

\item{}
\mdline{176}Lightweight.  We want the refinement features to be easy to use,
without the need for bulky syntax that reduces understanding.

\item{}
\mdline{179}Support reuse.  Though we have the one-developer view, we do wish
for constructs that lend themselves to reuse.

\item{}
\mdline{182}Modular verification and compilation.  We want it to be possible to
reason about a piece of code without having to know the details of
future refinements.  Similarly, we want it to be possible to compile
uses of an abstraction before all the details of the refinement have
been decided.
\end{itemize}

\section{\mdline{189}2.\hspace*{0.5em}\mdline{189}Refinement in Dafny}\label{sec-refinement-in-dafny}

\noindent\mdline{191}Dafny is a programming language designed with reasoning in mind\mdline{191}~[\mdcite{leino:dafny:lpar16}{16}]\mdline{191}.  Its
features include a repertoire of imperative and functional features.
In addition, the language integrates constructs for specifying the
intended behavior of programs, like pre- and postconditions, as
well as features that facilitate stating lemmas and writing proofs.
Dafny has a program verifier that checks that a program meets its
given specifications.  The integrated development environment (IDE)
for the language constantly runs the verifier in the background in
order to expedite feedback to the user\mdline{199}~[\mdcite{leinowuestholz:dafnyide}{19}]\mdline{199}.

\mdline{201}Dafny\mdline{201}'\mdline{201}s focus on reasoning and correctness makes it especially
appealing as a testbed for introducing refinement features.  We describe
Dafny\mdline{203}'\mdline{203}s refinement features in modules, in specifications of functions
and methods, in method bodies, and across modules.
In Section\mdline{205}~\mdref{sec-classes}{3}\mdline{205}, we use a longer example to describe
refinement features in classes.

\subsection{\mdline{208}2.0.\hspace*{0.5em}\mdline{208}Modules}\label{sec-modules}

\noindent\mdline{210}A Dafny program is divided into \mdline{210}\emph{modules}\mdline{210}.  A module contains
declarations of methods, functions, types (like inductive datatypes
and instantiable classes, where classes themselves declare fields, methods, and
functions), iterators, and nested modules.  In addition, a module can \mdline{213}\emph{import}\mdline{213}
other modules.

\mdline{216}One module can be declared to be a \mdline{216}\emph{refinement}\mdline{216} of another module, as
indicated by following the name of the new module with the keyword
\mdline{218}\mdcode{{\mdcolor{navy}refines}}\mdline{218} and the name of the module to be refined.
The refining module is based on the module it refines, but it is a
separate module.  More precisely, the contents of the refined module
is copied into the refining module, modulated by three kinds of
\mdline{222}\emph{directives}\mdline{222}:\mdline{222}\mdfootnote{0}{
\noindent\mdline{235}We speak about different \mdline{235}\emph{kinds}\mdline{235} of directives only
in order to explain the functionality provided in Dafny.  The user
never needs to name these directives when writing the program or
running the verifier or compiler.  Instead, which kind of
directive to apply is implicit from the program text, as we shall
see in examples.
\label{fn-fn-directives}
}\mdline{222}

\begin{itemize}

\item{}
\mdline{224}\textbf{Extend}\mdline{224} the refining module with additional declarations (for
example, declare a new type or a new method)

\item{}
\mdline{227}\textbf{Define}\mdline{227} entities whose definition the refined module omitted (for
example, define a previously opaque type or give a body of a
previously body-less function)

\item{}
\mdline{231}\textbf{Refine}\mdline{231} previously given specifications (for example,
strengthening postconditions) and previously given bodies of methods
and (in one special case) functions
\end{itemize}

\noindent\mdline{242}For example, Figure\mdline{242}~\mdref{fig-example}{\mdcaptionlabel{0}}\mdline{242} shows a module \mdline{242}\mdcode{A}\mdline{242} that declares an
opaque type \mdline{243}\mdcode{T}\mdline{243} and two functions, one of which (\mdline{243}\mdcode{F}\mdline{243}) is body-less
(that is, uninterpreted).  In this example, \mdline{244}\mdcode{A}\mdline{244} has been declared as
\mdline{245}\emph{abstract}\mdline{245}, which tells the compiler not to generate any code for \mdline{245}\mdcode{A}\mdline{245}.
Without the \mdline{246}\mdcode{{\mdcolor{navy}abstract}}\mdline{246} keyword, the compiler would complain about the
missing type definition and function body.  (Note, a module that
defines all its entities need not be abstract to be refined.)  Module
\mdline{249}\mdcode{B}\mdline{249} is declared as a refinement of \mdline{249}\mdcode{A}\mdline{249}.  It extends \mdline{249}\mdcode{A}\mdline{249} by declaring
an inductive datatype \mdline{250}\mdcode{T'}\mdline{250}.  It also defines \mdline{250}\mdcode{T}\mdline{250} to be a synonym for
\mdline{251}\mdcode{T'}\mdline{251} and it defines a body for \mdline{251}\mdcode{F}\mdline{251}.  Note that module \mdline{251}\mdcode{B}\mdline{251} also contains
function \mdline{252}\mdcode{Twice}\mdline{252}, which is copied from module \mdline{252}\mdcode{A}\mdline{252}.  Also, recall that the
presence of module \mdline{253}\mdcode{B}\mdline{253} in the program does not affect module \mdline{253}\mdcode{A}\mdline{253}; they
are two separate modules.

\begin{figure}[tbp]
\begin{mdcenter}
\begin{mdpre}
\noindent{\mdcolor{navy}abstract}~{\mdcolor{navy}module}~A~\{\\
~~{\mdcolor{navy}type}~T\\
~~{\mdcolor{navy}function}~F(x:~T):~T\\
~~{\mdcolor{navy}function}~Twice(x:~T):~T\\
~~\{~F(F(x))~\}\\
\}\\
{\mdcolor{navy}module}~B~{\mdcolor{navy}refines}~A~\{\\
~~{\mdcolor{navy}type}~T~=~T'\\
~~{\mdcolor{navy}datatype}~T'~=~Leaf({\mdcolor{navy}int})~\textbar{}~Node(T,~T)\\
~~{\mdcolor{navy}function}~F...\\
~~\{~{\mdcolor{navy}match}~x\\
~~~~{\mdcolor{navy}case}~Leaf(w)~=\textgreater{}~Leaf(w+{\mdcolor{purple}1})\\
~~~~{\mdcolor{navy}case}~Node(left,~right)~=\textgreater{}~Node(F(left),~F(right))\\
\}~\}
\end{mdpre}
\mdhr{}

\noindent\mdline{274}\mdcaption{\textbf{Figure~\mdcaptionlabel{0}.} \mdcaptiontext{Two example modules, one (\mdcode{B}) declared as a refinement of the other (\mdcode{A}).}}
\end{mdcenter}\label{fig-example}
\end{figure}

\mdline{275}In our example, we chose not to repeat the signature of \mdline{275}\mdcode{F}\mdline{275}, but
instead to use the syntax \mdline{276}\mdcode{...}\mdline{276}.  Dafny also allows the type signature
of \mdline{277}\mdcode{F}\mdline{277} to be repeated (allowing renamings of parameters) in the
refining module.\mdline{278}\mdfootnote{1}{
\noindent\mdline{280}We have considered requiring the \mdline{280}\mdcode{...}\mdline{280} syntax.  This
would always make it clear that the function is a refinement, and
it would reduce the clutter and brittle nature of having to
textually copy the signature.  However, as even this simple
example shows, the fact that the \mdline{284}\mdcode{...}\mdline{284} syntax does not repeat the
names of the parameters can also be confusing when looking at the
body of the function (\mdline{286}\textquotedblleft{}What is \mdcode{x}?\textquotedblright{}\mdline{286}).
\label{fn-fn-ellipsis}
}\mdline{278}

\mdline{288}Because Dafny\mdline{288}'\mdline{288}s refinement operates at the level of modules, it is
possible to simultaneously refine a set of types.  Compare this to the
limited one-type refinements achievable by a disciplined use of
subclassing in object-oriented languages.

\mdline{293}One mechanical way to describe the refinement features in Dafny is to
think of them as an elaborate template mechanism.  However, Dafny
restricts the use of the features to adhere to the standard \mdline{295}\emph{Principle of
Semantic Refinement}\mdline{296}, meaning that any client that is correct when
using a module \mdline{297}\mdcode{A}\mdline{297} is also guaranteed to be correct if \mdline{297}\mdcode{A}\mdline{297} is replaced
by any refinement of \mdline{298}\mdcode{A}\mdline{298}.  By analogy, object-oriented languages tend
to provide a syntactic mechanism for subclassing, but do not insist
that this mechanism be used only in accordance with \mdline{300}\emph{behavioral
subtyping}\mdline{301}~[\mdcite{dharaleavens:forcing}{5}, \mdcite{liskovwing94}{21}]\mdline{301}.  Since Dafny is
equipped with a program verifier, its definition can afford to insist
on following the Principle of Semantic Refinement (as opposed to just
providing a syntactic template mechanism).

\mdline{306}Next, we will start to see how Dafny\mdline{306}'\mdline{306}s restrictions preserve semantic
refinements.

\subsection{\mdline{309}2.1.\hspace*{0.5em}\mdline{309}Specifications}\label{sec-specifications}

\noindent\mdline{311}Dafny distinguishes between \mdline{311}\emph{methods}\mdline{311}, which are procedures with
statements that can modify the program\mdline{312}'\mdline{312}s heap, and \mdline{312}\emph{functions}\mdline{312}, which
are mathematical functions.  Both can have specifications:  pre-
and postconditions (given by \mdline{314}\mdcode{{\mdcolor{purple}requires}}\mdline{314} and \mdline{314}\mdcode{{\mdcolor{purple}ensures}~clauses}\mdline{314}), frame
specifications (\mdline{315}\mdcode{{\mdcolor{purple}modifies}}\mdline{315} clauses for methods and \mdline{315}\mdcode{{\mdcolor{purple}reads}}\mdline{315} clauses for
functions), and termination metrics (\mdline{316}\mdcode{{\mdcolor{purple}decreases}}\mdline{316} clauses).  A
refinement module is allowed to add more \mdline{317}\mdcode{{\mdcolor{purple}ensures}}\mdline{317} clauses, thus
strengthening the postcondition of the method or function.  In an
analogous way, it would be sound to weaken preconditions and shrink
frame specifications, but Dafny does not provide any syntax for doing
so.  Methods are allowed to be declared with \mdline{321}\mdcode{{\mdcolor{purple}decreases}~*}\mdline{321}, which says
that the method is allowed to diverge.  A refinement module is allowed
to change this specification by giving a termination metric
that proves termination.

\mdline{326}For example, method \mdline{326}\mdcode{Max}\mdline{326} in module \mdline{326}\mdcode{A}\mdline{326} of Figure\mdline{326}~\mdref{fig-spec}{\mdcaptionlabel{1}}\mdline{326} has a
weak specification.  It allows the method to diverge, and if the
method does terminate, the specification only says that the result
(which is returned in the output parameter \mdline{329}\mdcode{m}\mdline{329}) must not be smaller
than the input parameters.  Module 
\mdline{331}\mdcode{B}\mdline{331} strengthens the postcondition of \mdline{331}\mdcode{Max}\mdline{331} to say that the result is
one of the input parameters.  By giving a termination metric, it also
says that \mdline{333}\mdcode{Max}\mdline{333} terminates.

\mdline{335}Because Dafny enforces the Principle of Semantic Refinement, the
work of the verifier does not need to be repeated in refinement
modules.  In this example, when Dafny verifies module \mdline{337}\mdcode{A}\mdline{337}, it checks
that the implementation of \mdline{338}\mdcode{Max}\mdline{338} meets the weak postcondition.  When
it verifies module \mdline{339}\mdcode{B}\mdline{339}, it only checks that the implementation meets
the additional postcondition and that the lexicographic tuple
\mdline{341}\mdcode{x~\textless{}~y,~x~-~y}\mdline{341} strictly decreases with each recursive call.

\mdline{343}From the specification of \mdline{343}\mdcode{Max}\mdline{343} in \mdline{343}\mdcode{B}\mdline{343}, Dafny also verifies the
correctness of the assert statement in \mdline{344}\mdcode{Main}\mdline{344}.  Note, if no
termination metric is given for \mdline{345}\mdcode{Max}\mdline{345} in \mdline{345}\mdcode{B}\mdline{345}, then it would inherit
the \mdline{346}\textquotedblleft{}divergence allowed\textquotedblright{}\mdline{346} from \mdline{346}\mdcode{A}\mdline{346}; in that case, Dafny would complain
that \mdline{347}\mdcode{Main}\mdline{347}, which is not specified to allow divergence, is calling a
possibly diverging method.

\begin{figure}[tbp]
\begin{mdcenter}
\begin{mdpre}
\noindent{\mdcolor{navy}module}~A~\{\\
~~{\mdcolor{navy}method}~Max(x:~{\mdcolor{navy}int},~y:~{\mdcolor{navy}int})~{\mdcolor{navy}returns}~(m:~{\mdcolor{navy}int})\\
~~~~{\mdcolor{purple}ensures}~x~\textless{}=~m~\&\&~y~\textless{}=~m\\
~~~~{\mdcolor{purple}decreases}~*\\
~~\{\\
~~~~{\mdcolor{navy}if}~x~==~y~\{\\
~~~~~~m~{\mdcolor{navy}:=}~x;\\
~~~~\}~{\mdcolor{navy}else}~{\mdcolor{navy}if}~x~\textless{}~y~\{\\
~~~~~~m~{\mdcolor{navy}:=}~Max(y,~x);\\
~~~~\}~{\mdcolor{navy}else}~\{\\
~~~~~~m~{\mdcolor{navy}:=}~Max(x-{\mdcolor{purple}1},~y);\\
~~~~~~m~{\mdcolor{navy}:=}~m~+~{\mdcolor{purple}1};\\
\}~\}~\}\\
{\mdcolor{navy}module}~B~{\mdcolor{navy}refines}~A~\{\\
~~{\mdcolor{navy}method}~Max...\\
~~~~{\mdcolor{purple}ensures}~m~==~y~\textbar{}\textbar{}~m~==~x\\
~~~~{\mdcolor{purple}decreases}~x~\textless{}~y,~x~-~y\\
~~{\mdcolor{navy}method}~Main()~\{\\
~~~~{\mdcolor{navy}var}~m~{\mdcolor{navy}:=}~Max({\mdcolor{purple}10},~{\mdcolor{purple}20});\\
~~~~{\mdcolor{navy}assert}~m~==~{\mdcolor{purple}20};\\
\}~\}
\end{mdpre}
\mdhr{}

\noindent\mdline{375}\mdcaption{\textbf{Figure~\mdcaptionlabel{1}.} \mdcaptiontext{A convoluted implementation for computing the maximum of two numbers.  The specification of \mdcode{Max} in module \mdcode{B} strengthens the specification of \mdcode{Max} in \mdcode{A}.}}
\end{mdcenter}\label{fig-spec}
\end{figure}

\mdline{376}The two examples given so far show the directives Extend and Define.
The Refine directive is more involved, as we describe next.

\subsection{\mdline{379}2.2.\hspace*{0.5em}\mdline{379}Statements}\label{sec-statements}

\noindent\mdline{381}In what we have shown so far, a refining method can supply a body if
the refined method omitted it.  Dafny\mdline{382}'\mdline{382}s Refine directive goes
deeper than this and admits two kinds of change directives to a given
method body:

\begin{itemize}

\item{}
\mdline{386}\textbf{Tighten Up}\mdline{386} statements, to reduce nondeterminism

\item{}
\mdline{388}\textbf{Superimpose}\mdline{388} statements onto the refined method body, to
introduce and modify additional program state
\end{itemize}

\noindent\mdline{391}Since these directives apply to previously given statements or program
points, there is a need to explain, as part of the program, where the
directives are to apply.  For this purpose, we have borrowed the
\mdline{394}\emph{code skeletons}\mdline{394} from Chalice\mdline{394}~[\mdcite{leinoyessenov:chalicerefinement}{20}]\mdline{394}.
Code skeletons work by listing in the refining method the changes from
the refined method, when necessary mimicking the structure of the code
in the refined method.  We explain this functionality by example; see
\mdline{398}[\mdcite{leinoyessenov:chalicerefinement}{20}]\mdline{398} for a full merge algorithm.

\mdline{400}Dafny offers several nondeterministic statements.  These can be
replaced by more deterministic statements.  The replacement itself may
incur some proof obligation, but previous proof obligations are not
re-verified.  For example, the \mdline{403}\textquotedblleft{}assign such that\textquotedblright{}\mdline{403} statement \mdline{403}\mdcode{x~{\mdcolor{navy}:\textbar{}}~P;}\mdline{403}
says to set variable \mdline{404}\mdcode{x}\mdline{404} to any value satisfying the predicate \mdline{404}\mdcode{P}\mdline{404}
(there is a proof obligation that such an \mdline{405}\mdcode{x}\mdline{405} exists)\mdline{405}~[\mdcite{leino:epsilon}{18}]\mdline{405}.
By the Tighten Up
directive, this statement can be replaced by an ordinary
assignment statement \mdline{408}\mdcode{x~{\mdcolor{navy}:=}~E;}\mdline{408}, incurring a proof obligation that \mdline{408}\mdcode{P}\mdline{408}
with \mdline{409}\mdcode{x}\mdline{409} replaced by \mdline{409}\mdcode{E}\mdline{409} holds.

\mdline{411}For example, the pivot selection in QuickSort can first be implemented
by a statement
\begin{mdpre}
\noindent{\mdcolor{navy}var}~pivot~{\mdcolor{navy}:\textbar{}}~lo~\textless{}=~pivot~\textless{}~hi;
\end{mdpre}\noindent\mdline{416}and later refined to
\begin{mdpre}
\noindent{\mdcolor{navy}var}~p0,~p1,~p2~{\mdcolor{navy}:=}~lo,~(lo~+~hi)~/~{\mdcolor{purple}2},~hi~-~{\mdcolor{purple}1};\\
{\mdcolor{navy}if}~a[p2]~\textless{}~a[p0]~\{\\
~~p0,~p2~{\mdcolor{navy}:=}~p2,~p0;\\
\}\\
{\mdcolor{navy}var}~pivot~{\mdcolor{navy}:=}~{\mdcolor{navy}if}~a[p1]~\textless{}~a[p0]~{\mdcolor{navy}then}~p0~{\mdcolor{navy}else}~{\mdcolor{navy}if}~a[p2]~\textless{}~a[p1]~{\mdcolor{navy}then}~p2~{\mdcolor{navy}else}~p1;
\end{mdpre}\noindent\mdline{426}This refinement superimposes statements that declare and assign to new
local variables \mdline{427}\mdcode{p0}\mdline{427}, \mdline{427}\mdcode{p1}\mdline{427}, and \mdline{427}\mdcode{p2}\mdline{427}, and then tightens up the
assign-such-that statement to set \mdline{428}\mdcode{pivot}\mdline{428} according to the \mdline{428}\textquotedblleft{}median of
three\textquotedblright{}\mdline{429} strategy.  Dafny is able to distinguish
the superimposition from
the tighten up, since the merge algorithm matches the two
assignments\mdline{432}\textemdash{}\mdline{432}one nondeterministic in the refined module and one
deterministic in the refining module\mdline{433}\textemdash{}\mdline{433}to \mdline{433}\mdcode{pivot}\mdline{433}.
The refining module incurs a proof obligation that
the value it assigns to \mdline{435}\mdcode{pivot}\mdline{435} does indeed satisfy the condition
indicated in the refined module.

\mdline{438}The refining method is allowed to tighten up previous assignments and
to modify superimposed state, but is not otherwise allowed to assign
to previously declared variables.  We refer to this as the New State
Principle.  For instance, the assignments to the new local variable
\mdline{442}\mdcode{p0}\mdline{442} in the example above are allowed and so is the assignment that
tightens up the value of \mdline{443}\mdcode{pivot}\mdline{443}, but \mdline{443}\mdcode{pivot}\mdline{443} itself cannot be used as
a temporary variable to hold any intermediate values.

\mdline{446}Figure\mdline{446}~\mdref{fig-abs}{\mdcaptionlabel{2}}\mdline{446} shows another example where method \mdline{446}\mdcode{Abs}\mdline{446} is
specified to compute the absolute value of a given integer.  Module
\mdline{448}\mdcode{M0}\mdline{448} uses a nondeterministic \mdline{448}\mdcode{{\mdcolor{navy}if}}\mdline{448} statement that defines two control
paths.  One path sets the output parameter \mdline{449}\mdcode{a}\mdline{449} to \mdline{449}\mdcode{x}\mdline{449} and the other
hopes to make \mdline{450}\mdcode{a}\mdline{450} equal to \mdline{450}\mdcode{-x}\mdline{450} using a loop.  The method
implementation establishes the postcondition only if the assumed
conditions hold at the program points indicated.  Note, for example,
how the final assumption implies the last two conjuncts of the
postcondition.  Neither of the two \mdline{454}\mdcode{{\mdcolor{navy}assume}}\mdline{454} statements is provable in
module \mdline{455}\mdcode{M0}\mdline{455}; not the first, because not enough information is known
about \mdline{456}\mdcode{a}\mdline{456} after the loop, and not the second, because the \mdline{456}\mdcode{{\mdcolor{navy}if}}\mdline{456}
statement allows control to flow through either branch.

\mdline{459}Module \mdline{459}\mdcode{M1}\mdline{459} in Figure\mdline{459}~\mdref{fig-abs}{\mdcaptionlabel{2}}\mdline{459} refines \mdline{459}\mdcode{M0}\mdline{459} and tightens up the
choice of which \mdline{460}\mdcode{{\mdcolor{navy}if}}\mdline{460} branch to take.  This allows the second \mdline{460}\mdcode{{\mdcolor{navy}assume}}\mdline{460}
statement to be turned into an \mdline{461}\mdcode{{\mdcolor{navy}assert}}\mdline{461} statement.  That is, the
replacement of the \mdline{462}\mdcode{{\mdcolor{navy}assume}}\mdline{462} with an \mdline{462}\mdcode{{\mdcolor{navy}assert}}\mdline{462} incurs a proof obligation
that the condition does hold at that program point, which is provable
in module \mdline{464}\mdcode{M1}\mdline{464}.  The \mdline{464}\emph{elision statement}\mdline{464}, \mdline{464}\mdcode{...;}\mdline{464}, directs the
merge algorithm to match any code sequence.  Dafny implicitly
inserts an elision statement at the end of every code block, that is,
just before every \mdline{467}\textquotedblleft{}\mdcode{\}}\textquotedblright{}\mdline{467}, so all \mdline{467}\textquotedblleft{}\mdcode{...;}\textquotedblright{}\mdline{467} statements in the figure
could have been omitted.

\mdline{470}Dafny allows any number of refinement steps.  The figure shows module
\mdline{471}\mdcode{M1}\mdline{471} being further refined by module \mdline{471}\mdcode{M2}\mdline{471}.  It turns the first
\mdline{472}\mdcode{{\mdcolor{navy}assume}}\mdline{472} statement into an \mdline{472}\mdcode{{\mdcolor{navy}assert}}\mdline{472}, which is provable because of the
added loop invariant.  Note how expressions from the refined method
are not repeated but instead replaced by \mdline{474}\textquotedblleft{}\mdcode{...}\textquotedblright{}\mdline{474}.

\begin{figure}[tbp]
\begin{mdcenter}
\begin{mdpre}
\noindent{\mdcolor{navy}abstract}~{\mdcolor{navy}module}~M0~\{\\
~~{\mdcolor{navy}method}~Abs(x:~{\mdcolor{navy}int})~{\mdcolor{navy}returns}~(a:~{\mdcolor{navy}int})\\
~~~~{\mdcolor{purple}ensures}~(a~==~x~\textbar{}\textbar{}~a~==~-x)~\&\&~x~\textless{}=~a~\&\&~-x~\textless{}=~a\\
~~\{\\
~~~~{\mdcolor{navy}if}~*~\{\\
~~~~~~a~{\mdcolor{navy}:=}~x;\\
~~~~\}~{\mdcolor{navy}else}~\{\\
~~~~~~a~{\mdcolor{navy}:=}~{\mdcolor{purple}0};\\
~~~~~~{\mdcolor{navy}var}~b~{\mdcolor{navy}:=}~x;\\
~~~~~~{\mdcolor{navy}while}~b~\textless{}~{\mdcolor{purple}0}~\{\\
~~~~~~~~a,~b~{\mdcolor{navy}:=}~a~+~{\mdcolor{purple}1},~b~+~{\mdcolor{purple}1};\\
~~~~~~\}\\
~~~~~~{\mdcolor{navy}assume}~a~==~-x;\\
~~~~\}\\
~~~~{\mdcolor{navy}assume}~x~\textless{}=~a~\&\&~-x~\textless{}=~a;\\
\}~\}\\
{\mdcolor{navy}abstract}~{\mdcolor{navy}module}~M1~{\mdcolor{navy}refines}~M0~\{\\
~~{\mdcolor{navy}method}~Abs...~\{\\
~~~~{\mdcolor{navy}if}~{\mdcolor{purple}0}~\textless{}=~x~\{\\
~~~~~~...;\\
~~~~\}~{\mdcolor{navy}else}~\{\\
~~~~~~...;\\
~~~~\}\\
~~~~{\mdcolor{navy}assert}~...;\\
\}~\}\\
{\mdcolor{navy}module}~M2~{\mdcolor{navy}refines}~M1~\{\\
~~{\mdcolor{navy}method}~Abs...~\{\\
~~~~{\mdcolor{navy}if}~...~\{\\
~~~~~~...;\\
~~~~\}~{\mdcolor{navy}else}~\{\\
~~~~~~...;\\
~~~~~~{\mdcolor{navy}while}~...\\
~~~~~~~~{\mdcolor{purple}invariant}~a~+~x~==~b~\textless{}=~{\mdcolor{purple}0}\\
~~~~~~\{~...;~\}\\
~~~~~~{\mdcolor{navy}assert}~...;\\
~~~~\}\\
~~~~...;\\
\}~\}
\end{mdpre}
\mdhr{}

\noindent\mdline{518}\mdcaption{\textbf{Figure~\mdcaptionlabel{2}.} \mdcaptiontext{An artificial example that shows several Tighten Up refinements.  The \mdcode{Abs} method in module \mdcode{M0} postpones some proof obligations by introducing \mdcode{{\mdcolor{navy}assume}} statements, and leaves some room for later deciding which \mdcode{{\mdcolor{navy}if}} branch to take.  Module \mdcode{M1} tightens up the control flow and module \mdcode{M2} fills in missing parts of the program's correctness argument.}}
\end{mdcenter}\label{fig-abs}
\end{figure}

\mdline{519}Dafny provides a few statement refinement directives in addition to
the ones we have shown by the example above.  The general idea, as we have
shown, is for the refining methods to mimic the structure of the
method being refined, using \mdline{522}\mdcode{...;}\mdline{522} to stand for elided code,
superimposing new statements, and giving replacement statements that
tighten up nondeterminism in the refined method.  Dafny allows
statements to be labeled (which outside of refinement is useful with \mdline{525}\mdcode{{\mdcolor{navy}break}}\mdline{525} statements).
Labels can be repeated in a refining method, which can occasionally be
helpful as an aid for the merge algorithm.

\mdline{529}With one exception, the refining method is not allowed to disrupt
previous control flow.  For example, the refining method is not
allowed to add \mdline{531}\mdcode{{\mdcolor{navy}break}}\mdline{531} statements that exit out a loop.  The one
exception is that new \mdline{532}\mdcode{{\mdcolor{navy}return}}\mdline{532} statements are allowed.  Dafny checks
that the method\mdline{533}'\mdline{533}s postcondition holds at those points in the refining
method.  This is useful, for example, if the refinement adds a cache
or algorithmic support that enables a fast path in the method
implementation.

\mdline{538}Dafny includes two statements for the sole purpose of supporting
refinements, the elision statement and the \mdline{539}\mdcode{{\mdcolor{navy}modify}}\mdline{539} statement.  The
latter has the form
\begin{mdpre}
\noindent{\mdcolor{navy}modify}~W~\{~Body~\}
\end{mdpre}\noindent\mdline{544}where \mdline{544}\mdcode{W}\mdline{544} is a frame specification (which, like in a \mdline{544}\mdcode{{\mdcolor{purple}modifies}}\mdline{544}
clause, says which heap locations may be modified, and \mdline{545}\mdcode{\{~Body~\}}\mdline{545} is a
block statement.  Dafny treats the statement as the given block
statement, but enforces that its heap modifications are in
accordance with the frame specification.  As we shall see in Section
\mdline{549}\mdref{sec-classes}{3}\mdline{549}, the body of the \mdline{549}\mdcode{{\mdcolor{navy}modify}}\mdline{549} statement can be postponed and defined in
a refining method.  If the body is omitted, the semantics
of the statement is that of causing any arbitrary change permitted by
the frame specification.

\subsection{\mdline{554}2.3.\hspace*{0.5em}\mdline{554}Clients}\label{sec-clients}

\noindent\mdline{556}As one would expect from a language with a module system, Dafny allows
a module to \mdline{557}\emph{import}\mdline{557} other modules.  This makes the declarations in
the imported modules available to the importing module (the \mdline{558}\emph{client}\mdline{558})
via qualified names.  Since a module refinement gives rise to a
separate module, an issue arises of how a client selects among the
available refinements.

\mdline{563}The basic import declaration has the form:
\begin{mdpre}
\noindent{\mdcolor{navy}import}~X~=~M
\end{mdpre}\noindent\mdline{567}where \mdline{567}\mdcode{M}\mdline{567} is the name of a module defined elsewhere and \mdline{567}\mdcode{X}\mdline{567} is a local
name introduced as the qualifier when referring to declarations inside
\mdline{569}\mdcode{M}\mdline{569}.  In the common case where one chooses a local name identical
to the name of the imported module, the import declaration is
abbreviated by just \mdline{571}\mdcode{{\mdcolor{navy}import}~M}\mdline{571}.  The module import relation in a
program must be acyclic.  Moreover, an abstract module can be imported
only by other abstract modules.

\mdline{575}Consider a module \mdline{575}\mdcode{A0}\mdline{575} and a refinement module \mdline{575}\mdcode{A1}\mdline{575} (for brevity, we
show the modules without contents here):
\begin{mdpre}
\noindent{\mdcolor{navy}module}~A0~\{~\}\\
{\mdcolor{navy}module}~A1~{\mdcolor{navy}refines}~A0~\{~\}
\end{mdpre}\noindent\mdline{581}A client module can choose to import either one of these by using
\mdline{582}\mdcode{{\mdcolor{navy}import}~A~=~A0}\mdline{582} or \mdline{582}\mdcode{{\mdcolor{navy}import}~A~=~A1}\mdline{582}.

\mdline{584}It is also possible to be less specific, by replacing the \mdline{584}\mdcode{=}\mdline{584} with an
\mdline{585}\mdcode{{\mdcolor{navy}as}}\mdline{585}.  The import declaration
\begin{mdpre}
\noindent{\mdcolor{navy}import}~A~{\mdcolor{navy}as}~A0
\end{mdpre}\noindent\mdline{589}says to use \mdline{589}\mdcode{A}\mdline{589} as a local name for \mdline{589}\emph{some}\mdline{589} module that \mdline{589}\emph{adheres}\mdline{589} to
\mdline{590}\mdcode{A0}\mdline{590}, that is, whose contents (method bodies excluded)
is a superset of the contents of \mdline{591}\mdcode{A0}\mdline{591}.  The eventual module imported
can be \mdline{592}\mdcode{A0}\mdline{592} itself, any refinement of \mdline{592}\mdcode{A0}\mdline{592}, or in fact any other
module that structurally is like \mdline{593}\mdcode{A0}\mdline{593} or a refinement thereof.\mdline{593}\mdfootnote{2}{
\noindent\mdline{595}It is also possible to combine the \mdline{595}\mdcode{{\mdcolor{navy}as}}\mdline{595} and \mdline{595}\mdcode{=}\mdline{595}
imports: the declaration \mdline{596}\mdcode{{\mdcolor{navy}import}~A~{\mdcolor{navy}as}~A0~{\mdcolor{navy}default}~A1}\mdline{596} is essentially
treated like \mdline{597}\mdcode{{\mdcolor{navy}import}~A~{\mdcolor{navy}as}~A0}\mdline{597} by the verifier and as \mdline{597}\mdcode{{\mdcolor{navy}import}~A~=~A1}\mdline{597}
by the compiler.
\label{fn-fn-default}
}\mdline{593}

\mdline{600}An \mdline{600}\textquotedblleft{}\mdcode{{\mdcolor{navy}as}}\textquotedblright{}\mdline{600} import in a module can be tightened up in a refinement
module, as illustrated by the following example:
\begin{mdpre}
\noindent{\mdcolor{navy}module}~B0~\{\\
~~{\mdcolor{navy}import}~A~{\mdcolor{navy}as}~A0\\
\}\\
{\mdcolor{navy}module}~B1~{\mdcolor{navy}refines}~B0~\{\\
~~{\mdcolor{navy}import}~A~=~A1\\
\}
\end{mdpre}\noindent\mdline{612}Dafny checks that \mdline{612}\mdcode{A1}\mdline{612} adheres to \mdline{612}\mdcode{A0}\mdline{612}, which if \mdline{612}\mdcode{A1}\mdline{612} is a
module that refines \mdline{613}\mdcode{A0}\mdline{613} is a trivial check.  If \mdline{613}\mdcode{B1}\mdline{613} wants to rely on
the declarations in \mdline{614}\mdcode{A1}\mdline{614}that were not in \mdline{614}\mdcode{A0}\mdline{614}, but anticipates a
further refinement of the imported module, then it can instead use an
\mdline{616}\textquotedblleft{}\mdcode{{\mdcolor{navy}as}}\textquotedblright{}\mdline{616} import.

\mdline{618}As an example, consider the modules in Figure\mdline{618}~\mdref{fig-sort0}{\mdcaptionlabel{3}}\mdline{618}.  Module
\mdline{619}\mdcode{TotalOrder}\mdline{619} defines a type \mdline{619}\mdcode{T}\mdline{619}, a relation \mdline{619}\mdcode{Below}\mdline{619} on that type, and
an unproved lemma (that is, an axiom) that states a property of
\mdline{621}\mdcode{Below}\mdline{621}.  We have omitted lemma declarations for other properties that
might also be useful.  Module \mdline{622}\mdcode{GenericSorting}\mdline{622} imports some module
like \mdline{623}\mdcode{TotalOrder}\mdline{623}.  This lets it define methods (omitted in the
figure) that sort values of type \mdline{624}\mdcode{O.T}\mdline{624} according to the order
\mdline{625}\mdcode{O.Below}\mdline{625}.

\begin{figure}[tbp]
\begin{mdcenter}
\begin{mdpre}
\noindent{\mdcolor{navy}abstract}~{\mdcolor{navy}module}~TotalOrder~\{\\
~~{\mdcolor{navy}type}~T\\
~~{\mdcolor{navy}predicate}~Below(x:~T,~y:~T)\\
~~{\mdcolor{navy}lemma}~Transitive(x:~T,~y:~T,~z:~T)\\
~~~~{\mdcolor{purple}requires}~Below(x,~y)~\&\&~Below(y,~z)\\
~~~~{\mdcolor{purple}ensures}~Below(x,~z)\\
~~{\mdcolor{darkgreen}//~other~properties~omitted~from~the~figure}\\
\}\\
{\mdcolor{navy}abstract}~{\mdcolor{navy}module}~GenericSorting~\{\\
~~{\mdcolor{navy}import}~O~{\mdcolor{navy}as}~TotalOrder\\
~~{\mdcolor{darkgreen}//~sorting~methods~omitted~from~the~figure}\\
\}
\end{mdpre}
\mdhr{}

\noindent\mdline{643}\mdcaption{\textbf{Figure~\mdcaptionlabel{3}.} \mdcaptiontext{A sketch of a module that defines an ordering on a type \mdcode{T}, and the import declaration of a module that makes use of that ordering.}}
\end{mdcenter}\label{fig-sort0}
\end{figure}

\mdline{644}Figure\mdline{644}~\mdref{fig-sort1}{\mdcaptionlabel{4}}\mdline{644} shows refinements of the modules in Figure
\mdline{645}\mdref{fig-sort0}{\mdcaptionlabel{3}}\mdline{645}.  In particular, module \mdline{645}\mdcode{IntOrder}\mdline{645} defines \mdline{645}\mdcode{T}\mdline{645} to be a
synonym for \mdline{646}\mdcode{{\mdcolor{navy}int}}\mdline{646}, defines \mdline{646}\mdcode{Below}\mdline{646} to be the less-or-equal ordering on
integers, and gives a (trivial) proof that the property \mdline{647}\mdcode{Transitive}\mdline{647}
holds.  Module \mdline{648}\mdcode{IntSorting}\mdline{648} refines \mdline{648}\mdcode{GenericSorting}\mdline{648} by tightening up
the import declaration.  Consequently, the refining module will contain
copies of the refined module\mdline{650}'\mdline{650}s methods, but specialized for integers.

\begin{figure}[tbp]
\begin{mdcenter}
\begin{mdpre}
\noindent{\mdcolor{navy}module}~IntOrder~{\mdcolor{navy}refines}~TotalOrder~\{\\
~~{\mdcolor{navy}type}~T~=~{\mdcolor{navy}int}\\
~~{\mdcolor{navy}predicate}~Below...~\{~x~\textless{}=~y~\}\\
~~{\mdcolor{navy}lemma}~Transitive...~\{~\}\\
~~{\mdcolor{darkgreen}//~proofs~of~other~properties~omitted~from~the~figure}\\
\}\\
{\mdcolor{navy}module}~IntSorting~{\mdcolor{navy}refines}~GenericSorting~\{\\
~~{\mdcolor{navy}import}~O~=~IntOrder\\
\}
\end{mdpre}
\mdhr{}

\noindent\mdline{665}\mdcaption{\textbf{Figure~\mdcaptionlabel{4}.} \mdcaptiontext{The modules of Figure~\mdref{fig-sort0}{\mdcaptionlabel{3}} specialized to integers.}}
\end{mdcenter}\label{fig-sort1}
\end{figure}

\mdline{666}Note that the features we discuss in this paper do not give rise to
dynamic dispatch (like the \mdline{667}\emph{traits}\mdline{667} feature in Dafny does
\mdline{668}[\mdcite{dafny:traits}{2}]\mdline{668}).  There is no relation between refinement modules
that can be exploited dynamically at run time.

\section{\mdline{671}3.\hspace*{0.5em}\mdline{671}Classes and Data Refinement}\label{sec-classes}

\noindent\mdline{673}An important part of giving a simple description of a program lies in
choosing variables with simple types.  For example, sets and maps are
often used, but details of how to represent such sets and maps are
not.  The systematic coordinate transformation from such abstract data
structures to more efficient ones is called \mdline{677}\emph{data refinement}\mdline{677} (among
many other sources, see
\mdline{679}[\mdcite{backvonwright:book}{3}, \mdcite{griesprins:encapsulation}{8}, \mdcite{griesvolpano:transform}{9}]\mdline{679}).
Getting data refinement to work in the presence of classes is
difficult, because of encapsulation
issues with references to dynamically allocated objects
\mdline{683}[\mdcite{filipovic:seplogicrefinement}{6}, \mdcite{leinoyessenov:chalicerefinement}{20}]\mdline{683}.

\mdline{685}To present a small example that gives brief taste of the essential
problem, consider the following class:
\begin{mdpre}
\noindent{\mdcolor{navy}class}~Interval~\{~{\mdcolor{navy}var}~width:~{\mdcolor{navy}int}~\}
\end{mdpre}\noindent\mdline{690}With appropriate refinement rules, it is known how such a data
structure can be refined into, say:
\begin{mdpre}
\noindent{\mdcolor{navy}class}~IntervalEndPoints~\{\\
~~{\mdcolor{navy}var}~start:~{\mdcolor{navy}int}\\
~~{\mdcolor{navy}var}~end:~{\mdcolor{navy}int}\\
\}
\end{mdpre}\noindent\mdline{700}where \mdline{700}\mdcode{width}\mdline{700} is represented as the difference \mdline{700}\mdcode{end~-~start}\mdline{700}.
In this case, the fields \mdline{701}\mdcode{start}\mdline{701} and \mdline{701}\mdcode{end}\mdline{701} are introduced in the
refinement, and thus by the New State Principle, these assignments to
\mdline{703}\mdcode{start}\mdline{703} and \mdline{703}\mdcode{end}\mdline{703} are allowed in the refinement

\mdline{705}As an alternative refinement that involves reuse of library
components, suppose a library contains a class \mdline{706}\mdcode{Cell}\mdline{706}:
\begin{mdpre}
\noindent{\mdcolor{navy}class}~Cell~\{~{\mdcolor{navy}var}~data:~{\mdcolor{navy}int}~\}
\end{mdpre}\noindent\mdline{710}We may now consider a refinement like this:
\begin{mdpre}
\noindent{\mdcolor{navy}class}~IntervalCell~\{\\
~~{\mdcolor{navy}var}~start:~Cell\\
~~{\mdcolor{navy}var}~end:~Cell\\
\}
\end{mdpre}\noindent\mdline{719}where \mdline{719}\mdcode{width}\mdline{719} is represented as \mdline{719}\mdcode{end.data~-~start.data}\mdline{719}.  However,
the soundness of this kind of refinement is much more involved.
First, although the fields \mdline{721}\mdcode{start}\mdline{721} and \mdline{721}\mdcode{end}\mdline{721} are introduced in the
refining class, the field \mdline{722}\mdcode{data}\mdline{722} was available already in the program
being refined, and thus the simple New State Principle does not apply.
Instead, allowing the refinement to modify the values of
\mdline{725}\mdcode{start.data}\mdline{725} and \mdline{725}\mdcode{end.data}\mdline{725} requires more elaborate refinement rules.
The intuition is that the particular objects referenced by \mdline{726}\mdcode{start}\mdline{726} and
\mdline{727}\mdcode{end}\mdline{727} were never allocated in the program being refined, so
\mdline{728}\mdcode{start.data}\mdline{728} and \mdline{728}\mdcode{end.data}\mdline{728} in effect constitute new state.  For more
information about this problem, along with solutions, see
\mdline{730}[\mdcite{filipovic:seplogicrefinement}{6}, \mdcite{leinoyessenov:chalicerefinement}{20}]\mdline{730}.

\mdline{732}Dafny uses idioms of \mdline{732}\emph{dynamic frames}\mdline{732} to specify behavior of the heap
\mdline{733}[\mdcite{kassios:fm2006}{13}, \mdcite{leino:dafny:mod2008}{15}, \mdcite{smansetal:vericool}{28}]\mdline{733}.  The basic idea is to
programmatically keep track of the set of individual objects that
as an aggregate provide the behavior of the abstract object.  This
\mdline{736}\emph{representation set}\mdline{736} is often stored in a field
\begin{mdpre}
\noindent{\mdcolor{navy}ghost}~{\mdcolor{navy}var}~Repr:~{\mdcolor{navy}set}\ensuremath{\langle}{\mdcolor{navy}object}\ensuremath{\rangle}
\end{mdpre}\noindent\mdline{740}The field is declared as \mdline{740}\emph{ghost}\mdline{740}, meaning it is used only for
reasoning about the program.  The compiler erases ghost code, so
at run time they appear only in spirit
\mdline{743}[\mdcite{spiritofghostcode}{7}, \mdcite{leino:dafny:lpar16}{16}]\mdline{743}.

\mdline{745}Dafny does not have any specific data refinement or \mdline{745}\emph{transform}\mdline{745}
constructs\mdline{746}~[\mdcite{griesprins:encapsulation}{8}, \mdcite{griesvolpano:transform}{9}]\mdline{746}, but
the combination of ghost code, superimposition, and a directive that
allows predicates to be strengthened gives the ability to introduce
data structures in stages.  We proceed by giving an example,
introduced in several stages.

\subsection{\mdline{752}3.0.\hspace*{0.5em}\mdline{752}A Counter Specification}\label{sec-a-counter-specification}

\noindent\mdline{754}In the first stage, we give a specification of a very simple class,
see Figure\mdline{755}~\mdref{fig-m0}{\mdcaptionlabel{5}}\mdline{755}.  Abstractly, the class represents a counter,
whose value is stored in ghost field \mdline{756}\mdcode{N}\mdline{756}.  The class also declares a
field \mdline{757}\mdcode{Repr}\mdline{757} as described above and a predicate \mdline{757}\mdcode{Valid()}\mdline{757} that holds
when the object is in its steady state.  That is, the body of
\mdline{759}\mdcode{Valid()}\mdline{759} (omitted in module \mdline{759}\mdcode{M0}\mdline{759}) is the \mdline{759}\emph{class invariant}\mdline{759} of
\mdline{760}\mdcode{Counter}\mdline{760}~[\mdcite{meyer:oop}{22}]\mdline{760}.  (We explain the keyword \mdline{760}\mdcode{{\mdcolor{navy}protected}}\mdline{760} in
Section\mdline{761}~\mdref{sec-m2}{3.2}\mdline{761}.)

\mdline{763}The class also declares a constructor and two methods.  The last
postcondition of each of these is the familiar specification.  The
other parts of the specifications are exactly the idiomatic Dafny
dynamic-frame specifications for a constructor, a mutating method, and
a query method, respectively.\mdline{767}\mdfootnote{3}{
\noindent\mdline{780}By marking a class with the \mdline{780}\mdcode{\{{\mdcolor{purple}:autocontracts}\}}\mdline{780}
attribute, a pre-pass of the Dafny verifier will fill in the
idiomatic parts of specifications automatically, thus reducing
clutter in the program text.
\label{fn-fn-autocontracts}
}\mdline{767}
The occurrences of \mdline{768}\mdcode{Valid()}\mdline{768} express that the class invariant holds on
all method boundaries.  The conjuncts that mention \mdline{769}\mdcode{{\mdcolor{navy}fresh}}\mdline{769} say that any
objects that the constructor or mutating method add to the
representation set are freshly allocated, which is important for
callers to know\mdline{772}~[\mdcite{leino:dafny:mod2008}{15}]\mdline{772}.  Finally, the \mdline{772}\mdcode{{\mdcolor{purple}modifies}}\mdline{772}
clauses say that the constructor is only allowed to modify the state
of the object being constructed (which for the purpose of these
specifications is treated as if it was allocated immediately before the
constructor is called) and that \mdline{776}\mdcode{Inc}\mdline{776} is allowed to modify the state
of any object in the set \mdline{777}\mdcode{Repr}\mdline{777}.  In addition, every constructor and
method is allowed to allocate new object and modify their state.

\begin{figure}[tbp]
\begin{mdcenter}
\begin{mdpre}
\noindent{\mdcolor{navy}abstract}~{\mdcolor{navy}module}~M0~\{\\
~~{\mdcolor{navy}class}~Counter~\{\\
~~~~{\mdcolor{navy}ghost}~{\mdcolor{navy}var}~N:~{\mdcolor{navy}int}\\
~~~~{\mdcolor{navy}ghost}~{\mdcolor{navy}var}~Repr:~{\mdcolor{navy}set}\ensuremath{\langle}{\mdcolor{navy}object}\ensuremath{\rangle}\\
~~~~{\mdcolor{navy}protected}~{\mdcolor{navy}predicate}~Valid()\\
~~~~~~{\mdcolor{purple}reads}~{\mdcolor{navy}this},~Repr\\
~~~~{\mdcolor{navy}constructor}~()\\
~~~~~~{\mdcolor{purple}modifies}~{\mdcolor{navy}this}\\
~~~~~~{\mdcolor{purple}ensures}~Valid()~\&\&~{\mdcolor{navy}fresh}(Repr~-~\{{\mdcolor{navy}this}\})\\
~~~~~~{\mdcolor{purple}ensures}~N~==~{\mdcolor{purple}0}\\
~~~~{\mdcolor{navy}method}~Inc()\\
~~~~~~{\mdcolor{purple}requires}~Valid()\\
~~~~~~{\mdcolor{purple}modifies}~Repr\\
~~~~~~{\mdcolor{purple}ensures}~Valid()~\&\&~{\mdcolor{navy}fresh}(Repr~-~{\mdcolor{navy}old}(Repr))\\
~~~~~~{\mdcolor{purple}ensures}~N~==~{\mdcolor{navy}old}(N)~+~{\mdcolor{purple}1}\\
~~~~{\mdcolor{navy}method}~Get()~{\mdcolor{navy}returns}~(n:~{\mdcolor{navy}int})\\
~~~~~~{\mdcolor{purple}requires}~Valid()\\
~~~~~~{\mdcolor{purple}ensures}~n~==~N\\
\}~\}
\end{mdpre}
\mdhr{}

\noindent\mdline{808}\mdcaption{\textbf{Figure~\mdcaptionlabel{5}.} \mdcaptiontext{A module that gives the standard, idiomatic dynamic-frames specification of a simple class.}}
\end{mdcenter}\label{fig-m0}
\end{figure}

\mdline{809}Module \mdline{809}\mdcode{M0}\mdline{809} gives a client\mdline{809}'\mdline{809}s view of the \mdline{809}\mdcode{Counter}\mdline{809} class.  The
refinements that follow give the implementation of the class.

\subsection{\mdline{812}3.1.\hspace*{0.5em}\mdline{812}Defining Bodies}\label{sec-defining-bodies}

\noindent\mdline{814}We now define the predicate, constructor, and methods by giving them
bodies, see Figure\mdline{815}~\mdref{fig-m1}{\mdcaptionlabel{6}}\mdline{815}.  By separating modules \mdline{815}\mdcode{M0}\mdline{815} and \mdline{815}\mdcode{M1}\mdline{815}, we
simply achieve what in a language like, say, Modula-3 would be done by
writing a module interface and a module implementation\mdline{817}~[\mdcite{nelson:spwm3}{24}]\mdline{817}.

\begin{figure}[tbp]
\begin{mdcenter}
\begin{mdpre}
\noindent{\mdcolor{navy}abstract}~{\mdcolor{navy}module}~M1~{\mdcolor{navy}refines}~M0~\{\\
~~{\mdcolor{navy}class}~Counter~\{\\
~~~~{\mdcolor{navy}protected}~{\mdcolor{navy}predicate}~Valid...~\{\\
~~~~~~{\mdcolor{navy}this}~{\mdcolor{navy}in}~Repr~\&\&~{\mdcolor{navy}null}~!{\mdcolor{navy}in}~Repr\\
~~~~\}\\
~~~~{\mdcolor{navy}constructor}~...~\{\\
~~~~~~{\mdcolor{navy}ghost}~{\mdcolor{navy}var}~repr:~{\mdcolor{navy}set}\ensuremath{\langle}{\mdcolor{navy}object}\ensuremath{\rangle}~{\mdcolor{navy}:\textbar{}}~{\mdcolor{navy}null}~!{\mdcolor{navy}in}~repr~\&\&~{\mdcolor{navy}fresh}(repr);\\
~~~~~~N,~Repr~{\mdcolor{navy}:=}~{\mdcolor{purple}0},~repr~+~\{{\mdcolor{navy}this}\};\\
~~~~\}\\
~~~~{\mdcolor{navy}method}~Inc...~\{\\
~~~~~~N~{\mdcolor{navy}:=}~N~+~{\mdcolor{purple}1};\\
~~~~~~{\mdcolor{navy}modify}~Repr~-~\{{\mdcolor{navy}this}\};\\
~~~~\}\\
~~~~{\mdcolor{navy}method}~Get...~\{\\
~~~~~~n~{\mdcolor{navy}:\textbar{}}~{\mdcolor{navy}assume}~n~==~N;\\
\}~\}~\}
\end{mdpre}
\mdhr{}

\noindent\mdline{839}\mdcaption{\textbf{Figure~\mdcaptionlabel{6}.} \mdcaptiontext{A refinement of the module in Figure~\mdref{fig-m0}{\mdcaptionlabel{5}}, containing a simple \mdcode{Counter} class.  Module \mdcode{M1} defines the bodies in terms of the ghost fields \mdcode{Repr} and \mdcode{N}.}}
\end{mdcenter}\label{fig-m1}
\end{figure}

\mdline{840}Predicate \mdline{840}\mdcode{Valid()}\mdline{840} says that the receiver is always part of the
representation set, and the \mdline{841}\mdcode{{\mdcolor{navy}null}}\mdline{841} reference is not.  The constructor
needs to add to \mdline{842}\mdcode{Repr}\mdline{842} all objects that are to be part of the object\mdline{842}'\mdline{842}s
initial representation.  The details of this set are determined in
further refinements.  The constructor body in \mdline{844}\mdcode{M1}\mdline{844} anticipates these
further additions by introducing a local variable \mdline{845}\mdcode{repr}\mdline{845}, which it
allows to contain any set of newly allocated objects.

\mdline{848}Similarly, method \mdline{848}\mdcode{Inc}\mdline{848} uses the \mdline{848}\mdcode{{\mdcolor{navy}modify}}\mdline{848} statement, anticipating that
further refinements will want to do state changes of any representation
object other than \mdline{850}\mdcode{{\mdcolor{navy}this}}\mdline{850}.  (Note that by the New State Principle, a
refinement can still modify fields of \mdline{851}\mdcode{{\mdcolor{navy}this}}\mdline{851}, provided those fields
are declared in the refinement module.)

\mdline{854}Method \mdline{854}\mdcode{Get}\mdline{854} sets output parameter \mdline{854}\mdcode{n}\mdline{854} to \mdline{854}\mdcode{N}\mdline{854}, but in a somewhat
roundabout way.  First, in order to allow refinements to change how
\mdline{856}\mdcode{n}\mdline{856} is computed, \mdline{856}\mdcode{Get}\mdline{856} uses an assign-such-that statement rather than
a more straightforward assignment statement \mdline{857}\mdcode{n~{\mdcolor{navy}:=}~N;}\mdline{857}.  Second, since \mdline{857}\mdcode{n}\mdline{857}
is not a ghost variable, the right-hand side of the assignment
ordinarily must not depend on ghost variables like \mdline{859}\mdcode{N}\mdline{859}.  Use of the
keyword \mdline{860}\mdcode{{\mdcolor{navy}assume}}\mdline{860} in the assign-such-that statement indicates to Dafny
that this statement is not intended to be compiled, so Dafny relaxes
the ordinary restriction on ghost dependencies.\mdline{862}\mdfootnote{4}{
\noindent\mdline{864}The fact that \mdline{864}\mdcode{{\mdcolor{navy}assume}}\mdline{864} has the desired effect
here is rather coincidental.  It would probably be better to
change Dafny to allow ghost variables in right-hand sides of
assign-such-that statements in abstract modules.
\label{fn-fn-assume-such-that}
}\mdline{862}

\subsection{\mdline{869}3.2.\hspace*{0.5em}\mdline{869}An Implementation}\label{sec-m2}

\noindent\mdline{871}We introduce a concrete implementation of the counter.  We assume
there is some \mdline{872}\mdcode{Library}\mdline{872} module with a \mdline{872}\mdcode{Cell}\mdline{872} class and use two
instances of this class.  The value of the counter, \mdline{873}\mdcode{N}\mdline{873}, is
represented as the difference between the \mdline{874}\mdcode{data}\mdline{874} field of these two
objects, see Figure\mdline{875}~\mdref{fig-m2}{\mdcaptionlabel{7}}\mdline{875}.

\begin{figure}[tbp]
\begin{mdcenter}
\begin{mdpre}
\noindent{\mdcolor{navy}module}~M2~{\mdcolor{navy}refines}~M1~\{\\
~~{\mdcolor{navy}import}~Library\\
~~{\mdcolor{navy}class}~Counter~\{\\
~~~~{\mdcolor{navy}var}~c:~Library.Cell\\
~~~~{\mdcolor{navy}var}~d:~Library.Cell\\
~~~~{\mdcolor{navy}protected}~{\mdcolor{navy}predicate}~Valid...~\{\\
~~~~~~c~{\mdcolor{navy}in}~Repr~\&\&~d~{\mdcolor{navy}in}~Repr~\&\&\\
~~~~~~c~\ensuremath{\neq}~d~\&\&\\
~~~~~~N~==~c.data~-~d.data\\
~~~~\}\\
~~~~{\mdcolor{navy}constructor}~...~\{\\
~~~~~~c~{\mdcolor{navy}:=}~{\mdcolor{navy}new}~Library.Cell({\mdcolor{purple}0});\\
~~~~~~d~{\mdcolor{navy}:=}~{\mdcolor{navy}new}~Library.Cell({\mdcolor{purple}0});\\
~~~~~~{\mdcolor{navy}ghost}~{\mdcolor{navy}var}~repr:~{\mdcolor{navy}set}\ensuremath{\langle}{\mdcolor{navy}object}\ensuremath{\rangle}~{\mdcolor{navy}:=}~\{c,d\};\\
~~~~\}\\
~~~~{\mdcolor{navy}method}~Inc...~\{\\
~~~~~~...;\\
~~~~~~{\mdcolor{navy}modify}~...~\{\\
~~~~~~~~c.data~{\mdcolor{navy}:=}~c.data~+~{\mdcolor{purple}1};\\
~~~~~~\}\\
~~~~\}\\
~~~~{\mdcolor{navy}method}~Get...~\{\\
~~~~~~n~{\mdcolor{navy}:=}~c.data~-~d.data;\\
\}~\}~\}
\end{mdpre}
\mdhr{}

\noindent\mdline{905}\mdcaption{\textbf{Figure~\mdcaptionlabel{7}.} \mdcaptiontext{A further refinement of the module that defines the \mdcode{Counter} class.  This refinement implements the counter in terms of two dynamically allocated \mdcode{Cell} objects.}}
\end{mdcenter}\label{fig-m2}
\end{figure}

\mdline{906}The class is extended with the declaration of new fields \mdline{906}\mdcode{c}\mdline{906} and \mdline{906}\mdcode{d}\mdline{906}.
By superimpositions, the constructor straightforwardly allocates two
\mdline{908}\mdcode{Cell}\mdline{908} objects and assigns these to \mdline{908}\mdcode{c}\mdline{908} and \mdline{908}\mdcode{d}\mdline{908}.  The
constructor then tightens up the value assigned to \mdline{909}\mdcode{repr}\mdline{909}.

\mdline{911}Method \mdline{911}\mdcode{Inc}\mdline{911} defines a body for the \mdline{911}\mdcode{{\mdcolor{navy}modify}}\mdline{911} statement and method
\mdline{912}\mdcode{Get}\mdline{912} tightens up the assignment to \mdline{912}\mdcode{n}\mdline{912} by assigning it a value
computed from non-ghost fields.  To discharge the proof obligation
that the \mdline{914}\mdcode{{\mdcolor{navy}modify}}\mdline{914} body modifies only what is allowed by the frame
specification and the proof obligation incurred by the tighten-up
directive, it is necessary to have a stronger class invariant.  In
particular, the former proof obligation requires \mdline{917}\mdcode{c~{\mdcolor{navy}in}~Repr}\mdline{917} and the
latter requires \mdline{918}\mdcode{N~==~c.data~-~d.data}\mdline{918}.  In addition, the
well-formedness checks for the statements introduced require \mdline{919}\mdcode{c}\mdline{919} and
\mdline{920}\mdcode{d}\mdline{920} to be non-\mdline{920}\mdcode{{\mdcolor{navy}null}}\mdline{920}.

\mdline{922}Strengthening the class invariant comes down to changing the
definition of \mdline{923}\mdcode{Valid()}\mdline{923} to a stronger predicate.  This is dicey,
because \mdline{924}\mdcode{Valid()}\mdline{924} appears in preconditions and it is not sound to
strengthen preconditions in general.  Inside the refining module, the
verifier can arrange to re-verify proof obligations that involve
establishing \mdline{927}\mdcode{Valid()}\mdline{927} or assuming \mdline{927}\mdcode{!Valid()}\mdline{927}.  But what about client
modules that were verified against the module being refined?  Such
verifications would also have to be redone, which means verification
would no longer be modular.  For this reason, Dafny allows a predicate
to be strengthened only if it is marked as \mdline{931}\mdcode{{\mdcolor{navy}protected}}\mdline{931}, which means
the predicate\mdline{932}'\mdline{932}s exact definition will never be revealed outside the
module.  Consequently, other modules cannot rely on the exact
definition of the predicate, and so they are insensitive to any
changes of it.

\mdline{937}The syntax for this Predicate Strengthening directive is the same as
that to Define a predicate.  In other words, if a refining module
gives a body for a predicate that already had a body, the effect is
that of changing the definition of the predicate to the conjunction of
the two bodies.  This is allowed only for predicates marked as
\mdline{942}\mdcode{{\mdcolor{navy}protected}}\mdline{942}.

\mdline{944}It is possible to continue refining \mdline{944}\mdcode{M2}\mdline{944} into a subsequent module with
more state, but doing so requires changes to \mdline{945}\mdcode{M2}\mdline{945} that let it
anticipate further
refinements.  For example, module \mdline{947}\mdcode{M2}\mdline{947} may need to introduce
another variable like \mdline{948}\mdcode{repr}\mdline{948} in the constructor and to superimpose
another \mdline{949}\mdcode{{\mdcolor{navy}modify}}\mdline{949} statement in method \mdline{949}\mdcode{Inc}\mdline{949}\textemdash{}\mdline{949}in the same way that \mdline{949}\mdcode{M1}\mdline{949}
anticipated the further refinements given by \mdline{950}\mdcode{M2}\mdline{950}.
Methodologically, the fact that further refinements may require
changes to the module to be refined is justified (and even considered
normal) by the one-developer view.  An analogous situation arises in
object-oriented programming, when a new subclass needs an existing
class to introduce dynamically dispatched calls to a new method.

\mdline{957}Rather than taking our example in the direction of adding more state,
we will in the next subsection illustrate the gist of a performance
optimization that does not require further data refinements.

\subsection{\mdline{961}3.3.\hspace*{0.5em}\mdline{961}A Performance Optimization}\label{sec-a-performance-optimization}

\noindent\mdline{963}Figure\mdline{963}~\mdref{fig-m3}{\mdcaptionlabel{8}}\mdline{963} shows a module \mdline{963}\mdcode{M3}\mdline{963} that refines \mdline{963}\mdcode{M2}\mdline{963}.  It applies a
directive only to method \mdline{964}\mdcode{Get}\mdline{964}, into whose body it superimposes an \mdline{964}\mdcode{{\mdcolor{navy}if}}\mdline{964}
statement.  The new code sets up a fast path in the event that the
stated condition holds.  In Dafny, the \mdline{966}\mdcode{{\mdcolor{navy}return}}\mdline{966} statement with
argument expressions has the effect of assigning the expressions to
the output parameters and then returning from the method.  Since
output parameters do not fall under the New State Principle, the
refinement is normally not allowed new assignments to them; however,
as this is a useful and harmless case, the implicit assignments to
output parameters that happens as part of a superimposed \mdline{972}\mdcode{{\mdcolor{navy}return}}\mdline{972}
statement are allowed.

\mdline{975}In our simple example, the fast path we introduced will not give rise
to any actual performance improvement unless the compiler realizes
that \mdline{977}\mdcode{d.data~==~{\mdcolor{purple}0}}\mdline{977} actually always holds (in which case the condition
does not need to be tested in the emitted code).  To illustrate how
refinements could help give that information to the compiler, we can
strengthen the class invariant further, see module \mdline{980}\mdcode{M4}\mdline{980} in Figure
\mdline{981}\mdref{fig-m3}{\mdcaptionlabel{8}}\mdline{981}.

\begin{figure}[tbp]
\begin{mdcenter}
\begin{mdpre}
\noindent{\mdcolor{navy}module}~M3~{\mdcolor{navy}refines}~M2~\{\\
~~{\mdcolor{navy}class}~Counter~\{\\
~~~~{\mdcolor{navy}method}~Get...~\{\\
~~~~~~{\mdcolor{navy}if}~d.data~==~{\mdcolor{purple}0}~\{~{\mdcolor{navy}return}~c.data;~\}\\
~~~~~~...;\\
\}~\}~\}\\
{\mdcolor{navy}module}~M4~{\mdcolor{navy}refines}~M3~\{\\
~~{\mdcolor{navy}class}~Counter~\{\\
~~~~{\mdcolor{navy}protected}~{\mdcolor{navy}predicate}~Valid...~\{~d.data~==~{\mdcolor{purple}0}~\}\\
\}~\}
\end{mdpre}
\mdhr{}

\noindent\mdline{997}\mdcaption{\textbf{Figure~\mdcaptionlabel{8}.} \mdcaptiontext{Module \mdcode{M3} adds a fast path to the \mdcode{Get} method of module \mdcode{M2}, and module \mdcode{M4} strengthens predicate \mdcode{Valid()} to demonstrate that \mdcode{d.data~==~{\mdcolor{purple}0}} is in fact an invariant of the class.}}
\end{mdcenter}\label{fig-m3}
\end{figure}

\mdline{998}It is worth mentioning once more that \mdline{998}\mdcode{M2}\mdline{998}, \mdline{998}\mdcode{M3}\mdline{998}, and \mdline{998}\mdcode{M4}\mdline{998} are three
separate modules.  Dafny checks the refinement among these successive
modules, but does not relate the classes they define.  In particular,
classes \mdline{1001}\mdcode{M2.Counter}\mdline{1001}, \mdline{1001}\mdcode{M3.Counter}\mdline{1001}, and \mdline{1001}\mdcode{M4.Counter}\mdline{1001} are three
separate types and are not subclasses of one another.

\mdline{1004}This completes our illustration of how a class can be built in
stages.  Looking back at Figures\mdline{1005}~\mdref{fig-m0}{\mdcaptionlabel{5}}\mdline{1005} through\mdline{1005}~\mdref{fig-m3}{\mdcaptionlabel{8}}\mdline{1005}, the
elisions are such that the refinements from module to module stand
out.  A user can inspect what any ellipsis stands for by placing the
mouse pointer above the ellipsis in the Dafny IDE, upon which the elided
information will be displayed as a hover text.

\section{\mdline{1011}4.\hspace*{0.5em}\mdline{1011}Experience, Evaluation, and Related Work}\label{sec-experience-evaluation-and-related-work}

\noindent\mdline{1013}We have used the refinement features in Dafny for a number of toy
programs.  Although the provided directives can accomplish the usual
refinement tasks, our impression is that refinement works more
smoothly on paper than in our language.  Things that, due to
hand waving, may be simple to achieve on paper (like the problem
solved by the local variable \mdline{1018}\mdcode{repr}\mdline{1018} in Figure\mdline{1018}~\mdref{fig-m1}{\mdcaptionlabel{6}}\mdline{1018}) look more
clumsy in our language design.

\mdline{1021}One could argue that useful formal-methods techniques are also useful
if applied informally, that is, without actually carrying through the
proofs.  This argument leads to asking if our superimposition and
tighten-up directives are useful devices for program structuring.
Here, too, it is not clear that our design gets a good score.  For
one, the fact that one needs to declare another module in order to
stage some refinements can feel bulky.

\mdline{1029}A similar bulkiness issue also arises in Event-B\mdline{1029}~[\mdcite{abrial:eventb:book}{0}]\mdline{1029} implemented in
the Rodin tool\mdline{1030}~[\mdcite{rodintoolset}{1}]\mdline{1030}, where all unchanged events have to be
copied into the file that contains the subsequent refinement.  An
alternative is given by the \mdline{1032}\textquotedblleft{}refinement layer\textquotedblright{}\mdline{1032} annotations in Civil
\mdline{1033}[\mdcite{civil:cav2015}{12}]\mdline{1033}.  These allow several stages of refinement to be
given in a single source text.  The verifier processes a given program
once for each declared layer, suitably ignoring the declarations of
higher-numbered layers.

\mdline{1038}When authoring or reading a sequence of refinements, one sometimes
wants to see only the changes from one module to the next and
sometimes wants to see the full resulting program.  Our elision
statements only address the former, and our IDE\mdline{1041}'\mdline{1041}s hover text does not
adequately address the latter.  We had chosen the elision statements
under a rather traditional view that a program is a printable piece of
program text.  A more modern or even futuristic view would be to let
the sequence of refinements appear as layered text in the IDE.  A
user could then be given various ways to input and read the program.
The refinement tools KIV\mdline{1047}~[\mdcite{kiv:overview}{26}]\mdline{1047} and Rodin\mdline{1047}~[\mdcite{rodintoolset}{1}]\mdline{1047}
have embraced the idea that the IDE
can manage the program better than a line-by-line editor can.  We
hope such environments will also be developed for languages that look
more similar to today\mdline{1051}'\mdline{1051}s mainstream languages than KIV and Event-B do.

\mdline{1053}A desirable scenario to support in staged program development is to
write a program in an abstract way and then replace the operations on
certain variables with other, more efficient operations on alternative
variables.  This is a central goal of the \mdline{1056}\emph{transform}\mdline{1056} by Gries et
al.\mdline{1057}~[\mdcite{griesprins:encapsulation}{8}, \mdcite{griesvolpano:transform}{9}]\mdline{1057}.  At first,
the rather syntactic match-and-replace rules in these transforms
appear brittle.  But given that this is a scenario we would like to
support smoothly, and given that we are buying into the
one-developer-view idea of
anticipating refinements, we would be interested in incorporating the
transform into Dafny.

\mdline{1065}The Dafny design that a refinement module creates a separate module is a
feature in some cases.  For example, it allows multiple refinements of
the \mdline{1067}\mdcode{TotalOrder}\mdline{1067} module in Figure\mdline{1067}~\mdref{fig-sort0}{\mdcaptionlabel{3}}\mdline{1067}, each one of which can
benefit from reuse.  It has also been used to define common processing
of different services in the IronFleet project, which was authored in
Dafny\mdline{1070}~[\mdcite{ironfleet:sosp2015}{11}]\mdline{1070}.
But we have also seen it make the common
interface-implementation pattern rather verbose, since it requires
a refinement module when a client wants to tighten up which
implementation gets used for the abstract module it \mdline{1074}\mdcode{{\mdcolor{navy}as}}\mdline{1074}-imported.
(We have started exploring an alternative module design wherein every
abstract module has a default refinement module.)

\mdline{1078}An early tool for machine-assisted program development
lets the user apply \mdline{1079}\emph{refinement tactics}\mdline{1079} to massage a formal specification
into code\mdline{1080}~[\mdcite{grovesnicksonutting:refinementtactics}{10}]\mdline{1080}.  The tactics applied are
recorded and can be displayed.  Moreover, the IDE allows a user to expand
a sub-specification to see what it has been refined into; conversely,
the details can be elided to instead show just the more abstract
sub-specification that they implement.  We would wish for an IDE that keeps
track of the program-derivation tree in this way.  However, we also note
that Dafny provides greater flexibility in introducing correlated transformations
(like the addition of an \mdline{1087}\mdcode{{\mdcolor{navy}assume}}\mdline{1087} statement in each branch of an \mdline{1087}\mdcode{{\mdcolor{navy}if}}\mdline{1087} to justify
some other refinement transformation after the \mdline{1088}\mdcode{{\mdcolor{navy}if}}\mdline{1088} statement), and it is
not clear how these can be presented with an equally simple IDE. \mdline{1089} \mdline{1089}

\mdline{1091}Despite many shortcomings in our language design, the current
refinement features in Dafny have been useful in some complex
examples.  One such example is a break-down of the Schorr-Waite
algorithm into stages.  More precisely, the proof obligations, loop
invariants, and ghost variables used in the proof were broken down
into a sequence of refinements that seems to separate concerns in a
desired way.  Another example is the formalization of the Cloudmake
algorithm\mdline{1098}~[\mdcite{dafny:cloudmake}{4}]\mdline{1098}.  It introduces some axiomatized
functions and later uses refinements to prove the feasibility of those
axioms.  Interestingly enough, these examples use the refinement
features mostly to structure \mdline{1101}\emph{proofs}\mdline{1101}, not to structure the executable
statements of the program.

\mdline{1104}The examples in our paper can be tried online at
\mdline{1105}\mdcode{http://rise4fun.com/Dafny/}\mdline{1105}\{\mdline{1105}\href{http://rise4fun.com/Dafny/4FH}{\mdcode{4FH}}\mdline{1105},
\mdline{1106}~\href{http://rise4fun.com/Dafny/74s9}{\mdcode{74s9}}\mdline{1106},
\mdline{1107}~\href{http://rise4fun.com/Dafny/jrJQ}{\mdcode{jrJQ}}\mdline{1107},
\mdline{1108}~\href{http://rise4fun.com/Dafny/jX5Y}{\mdcode{jX5Y}}\mdline{1108},
\mdline{1109}~\href{http://rise4fun.com/Dafny/n07}{\mdcode{n07}}\mdline{1109}\}.
Additional examples can be found in the Dafny test suite at
\mdline{1111}\href{http://dafny.codeplex.com}{\mdcode{http://dafny.codeplex.com}}\mdline{1111}.
A video of a SPLASH 2012 keynote with live demos is also available
online\mdline{1113}~[\mdcite{leino:splash2012:keynote}{17}]\mdline{1113}.

\section{\mdline{1137}5.\hspace*{0.5em}\mdline{1137}Concluding Remarks}\label{sec-concluding-remarks}

\noindent\mdline{1139}We have described the refinement features in version 1.9.5 of Dafny.
While far from perfect, we have combined refinement and automated
verification into a programming language.  We hope that use of our
system will inspire further exploration and innovation in
incorporating refinement features in day-to-day programming languages.

\subsection*{\mdline{1145}Acknowledgments}\label{sec-acknowledgments}

\noindent\mdline{1147}We are grateful to Lindsay Groves and Mark Utting for feedback at the
REFINE 2015 workshop, and to the referees for their detailed readings
and constructive feedback.

\mdsetrefname{References}
{\mdbibindent{0}
}

\end{document}